\documentclass[twocolumn,showpacs,preprintnumbers,amsmath,amssymb]{revtex4}

\usepackage{graphicx}% Include figure files
\usepackage{dcolumn}% Align table columns on decimal point
\usepackage{bm}% bold math
\usepackage{amsmath,epsfig}

\begin{document}

%\baselineskip=22pt

%\textbf{Classification}: PHYSICAL SCIENCE, Applied Physical Science\\

\title{Hierarchically nested factor model from multivariate data}

\author{M. Tumminello$^{(1)}$, F. Lillo$^{(1,2)}$ \& R. N. Mantegna$^{(1)}$ }

\affiliation{
$^{(1)}$ 
Dipartimento di Fisica e Tecnologie Relative, 
Universit\`a di Palermo - Viale delle Scienze, I-90128 Palermo, Italy\\
$^{(2)}$ Santa Fe Institute, 1399 Hyde Park Road, Santa Fe, NM 87501, U.S.A.
}

\date{\today}

\begin{abstract}
We show how to achieve a statistical description of the hierarchical structure of a multivariate data set. Specifically we show that the similarity matrix resulting from a hierarchical clustering procedure is the correlation matrix of a factor model, the hierarchically nested factor model. In this model, factors are mutually independent and hierarchically organized. Finally, we use a bootstrap based procedure to reduce the number of factors in the model with the aim of retaining only those factors significantly robust with respect to the statistical uncertainty due to the finite length of data records.
\end{abstract}

\maketitle

Many complex systems observed in the physical, biological and social sciences are organized in a nested hierarchical structure, i.e. the elements of the system can be partitioned in clusters which in turn can be partitioned in subclusters and so on up to a certain level \cite{Simon62,Anderson72}. 
Several examples of hierarchically organized physical \cite{Palmer1984,Sethna93}, biological \cite{Roskelley95,Barabasi02,Csete02} and social \cite{Malone94,Mantegna1999,Vespignani01,Tumminello05} systems have been investigated in the literature. The hierarchical structure of interactions among elements strongly affects the dynamics of complex systems. Therefore, a quantitative description of hierarchical properties of the system is a key step in the modeling of complex systems. In this letter, we address the problem of inferring a factor model from a multivariate data set. A factor model is a mathematical model which attempts to explain the correlation between a large set of variables in terms of a small number of underlying factors. A major assumption of factor analysis is that it is not possible to observe these factors directly; the variables depend upon the factors but are also subject to random errors \cite{Mardia}. We show that the factor model we introduce fully describes the hierarchical structure of interactions among elements of the complex system. Such a structure is elicited by hierarchical clustering of multivariate data. 
The analysis of multivariate data provides crucial
information in the investigation of a wide variety of systems.
Multivariate analysis methods are designed to extract information both
on the number of main factors characterizing the dynamics of the
investigated system and on the composition of the groups (clusters) 
in which the system is intrinsically organized. Recently, physicists started to contribute to the development of new multivariate techniques (e.g. \cite{Tumminello05, Blatt96, Hutt99, Giada01, Kraskov05, Tsafrir05, Slonim05}).
Among multivariate techniques, natural candidates for detecting the hierarchical structure of a set of data are hierarchical clustering methods \cite{Anderberg}. These methods allow to associate a dendrogram with a correlation matrix (or more generally with a similarity matrix), i.e. they give a schematic description of hierarchies. It is worth pointing out that the whole information contained in the dendrogram can be stored in a filtered similarity matrix ${\bf C^{<}}$ \cite{Anderberg}. The matrix ${\bf{C^<}}$ has well defined metric properties. When the matrix ${\bf{C^<}}$ of elements $\rho_{i j}^<$ is obtained by starting from a correlation matrix, then the matrix of distances $d_{i j}^<=\sqrt{2 (1-\rho_{i j}^<)}$ has ultrametric properties \cite{Rammal86}. 

In this letter, we answer the following scientific question: given a multivariate data set is it possible to construct a factor model retaining the whole information about hierarchies which is detected by a hierarchical clustering? In the following, we show that it is possible to give a description of hierarchies detected by hierarchical clustering in terms of a factor model, termed Hierarchically Nested Factor Model (HNFM). This model is constructed in such a way that its correlation matrix coincides with the similarity matrix ${\bf C^{<}}$ filtered by the chosen hierarchical clustering procedure.
Furthermore, for a hierarchical clustering performed by estimating a correlation matrix from an empirical data set which is unavoidably of finite size, i.e. a set of  $N$ elements each characterized by a number $T$ of records, we provide a bootstrap based methodology allowing to remove from the model those factors which are characterized by a statistical reliability smaller than a predefined standard threshold, e.g. 95\%. In this letter, we consider time series, however the results are general and also valid for any investigation of multivariate data.  
There are many clustering algorithms \cite{Anderberg}, here we use the Average Linkage Cluster Analysis (ALCA). However, we wish to point out that our technique can be used with most clustering algorithms giving a dendrogram \cite{note1}, such as, for example, the single linkage clustering algorithm.

Hereafter, we provide a methodology to associate a nested factor model with a multivariate data set. The association is done by retaining all the information about the hierarchies detected by a hierarchical clustering. This is achieved by considering a factor model in bijective relation with a dendrogram (or with the filtered matrix ${\bf C^<}$),which is the output of a hierarchical clustering. 
We are going to introduce our method by making use of the illustrative dendrogram given in Fig. \ref{dendro}.
\begin{figure}
\includegraphics[width=0.45\textwidth]{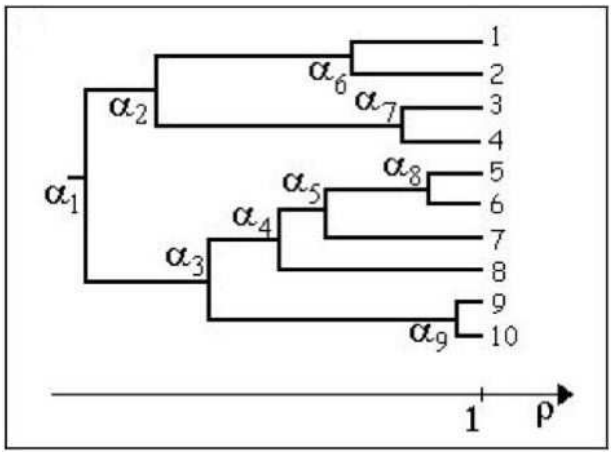}
\caption{ Illustrative example of a rooted tree associated with a system of $N=10$ elements (leaves in the tree). The symbols $\{\alpha_1,...,\alpha_9\}$ labels the $N-1=9$ internal nodes.} 
\label{dendro}
\end{figure}
A dendrogram is a rooted tree, i.e. a tree in which a special node (the root) is singled out. In our example this node is $\alpha_1$.  In the rooted tree, we distinguish between leaves and internal nodes. Specifically, vertices of degree $1$ are representing leaves (vertices labeled $1, 2,..., 10$ in Fig. \ref{dendro}) while vertices of degree greater than 1 are representing internal nodes (vertices labeled $\alpha_1$, $\alpha_2$,..., $\alpha_9$ in Fig. \ref{dendro}). 
We associate a \emph{genealogy} $G(i)$  ($G(\alpha_h)$) with each leaf $i$ (internal node $\alpha_h$).The genealogy is the ordered set of internal nodes connecting leaf $i$ (internal node $\alpha_h$) to the root $\alpha_1$. For instance, in Fig. \ref{dendro}, the genealogy associated with the leaf 3 is $G(3)=\{\alpha_7, \alpha_2, \alpha_1 \}$ and the genealogy of the internal node $\alpha_7$ is $G(\alpha_7)=\{\alpha_7, \alpha_2, \alpha_1 \}$. Note that the internal node $\alpha_7$ is included in $G(\alpha_7)$. Finally, we say that an internal node $w$ is the {\it  parent} of the node $v$, and we use the notation $w=g(v)$, if $w$ immediately precedes $v$ on the path from the root to $v$. For example it is $\alpha_2=g(\alpha_7)$ in Fig. \ref{dendro}.
Beside the topological structure, dendrograms obtained through standard hierarchical clustering algorithms applied to a correlation matrix have also metric properties. In fact, clustering algorithms associate a correlation coefficient  $\rho_{\alpha_i}$ with each internal node $\alpha_i$  \cite{Anderberg}. Our internal node labeling implies that  $\rho_{\alpha_i} \le \rho_{\alpha_{i+1}}$ and here we consider $\rho_{\alpha_1}\ge0$ \cite{note2}. 
The whole information about the rooted tree is stored in the $N \times N$ matrix ${\bf C^<}$ of elements $\rho_{ij}^<=\rho_{\alpha_k}$, where $\alpha_k$ is the first internal node in which leaves $i$ and $j$ are merged together \cite{Anderberg}. For example, in Fig. \ref{dendro}, it is $\rho_{3 7}^<=\rho_{\alpha_1}$ and $\rho_{5 7}^<=\rho_{\alpha_5}$.
In ${\bf C^{<}}$ there are at most $N-1$ distinct coefficients. Exactly $N-1$ distinct coefficients are obtained in case of binary rooted trees. Since any rooted tree can be obtained from a rooted binary tree by introducing a degeneracy of nodes, in the following we consider binary rooted trees. 

Here we show  that the matrix ${\bf C^<}$ is the correlation matrix of a HNFM defined as 
\begin{equation}
\label{model}
x_i (t)=\sum_{\alpha_h \in G(i)} {\gamma_{\alpha_h} f^{(\alpha_h)}(t)}+\eta_i\,\, \epsilon_i (t),
\end{equation}
where $i\in\{1,...,N\}$, $\eta_i=[1-\sum_{\alpha_h \in G(i)} {\gamma_{\alpha_h}^{2}}]^{1/2}$. The $h^{th}$ factor $f^{(\alpha_h)}(t)$ and $\epsilon_i (t)$ are independent identically distributed (i.i.d.) random variables with zero mean and unit variance.
In order to ensure that the correlation matrix of the model of Eq.~(\ref{model}) is ${\bf C^<}$, the $\gamma$ parameters need to be chosen as 
\begin{eqnarray}
\label{coeffic}
%\left \{  \right.
%\begin{aligned}
\gamma_{\alpha_1} = & \sqrt{\rho_{\alpha_1}} & \nonumber \\
\gamma_{\alpha_h} =  & \sqrt{\rho_{\alpha_h}-\rho_{g(\alpha_h)}} & \,\,\,\,\,\,\, \forall \, h=2,...,n-1
%\end{aligned}
%\right.
\end{eqnarray}
where, assuming $\rho_{\alpha_1}\ge0$, all the coefficients $\gamma_{\alpha_h}$ are non negative real numbers. Hereafter we show that the matrix ${\bf C^<}$ is the correlation matrix of the factor model of Eq. (\ref{model}) with coefficients $\gamma$'s given in Eq.~(\ref{coeffic}). Let us consider a generic pair of elements $i$ and $j$ merging together at the node $\alpha_k$ corresponding to the correlation level $\rho_{\alpha_k}$. We prove that the cross correlation $\langle x_i x_j \rangle$ equals the correlation $\rho_{i j}^< = \rho_{\alpha_k}$. In fact, the cross correlation $\langle x_i x_j \rangle$ depends only on the factors $f^{(\alpha_h)}$ which are common to $x_i$ and $x_j$. Since we associate a factor with each internal node, we need to identify the internal nodes belonging to both the genealogies $G(i)$ and $G(j)$. 
One can verify that  $G(i) \cap G(j)=G(\alpha_k)$. For example, in Fig. \ref{dendro} we have that  $G(2)=\{\alpha_6, \alpha_2, \alpha_1 \}$ and $G(3)=\{\alpha_7, \alpha_2, \alpha_1 \}$ so that $G(2)\cap G(3)=\{\alpha_2, \alpha_1 \}=G(\alpha_2)$. 
By making use of Eqs.~(\ref{model}, \ref{coeffic}) the cross correlation between variables $x_i$ and $x_j$ is
\begin{equation}
\label{scalar}
\left\langle x_i x_j \right\rangle = \sum_{\alpha_h \in G(\alpha_k)} {\gamma_{\alpha_h}^2} = \rho_{\alpha_k}=\rho_{i j}^<.%\\
\end{equation} 
For example, with reference to Fig.~\ref{dendro}, we have $\left\langle x_2 x_3 \right\rangle= \gamma_{\alpha_2}^2+\gamma_{\alpha_1}^2=\rho_{\alpha_2}-\rho_{\alpha_1}+\rho_{\alpha_1}=\rho_{\alpha_2}$.
Thus the matrix ${\bf C^<}$ is the correlation matrix associated with the factor model of Eq. (\ref{model}). It is worth noting that the  matrix ${\bf C^<}$ is positive definite, because, as we have shown,  ${\bf C^<}$ is the correlation matrix of a factor model. 
In conclusion, the HNFM is a factor model  taking into account the hierarchical properties of the investigated system which are elicited from data by hierarchical clustering.  

It is worth pointing out that the simple investigation of the eigenvectors of the correlation matrix is not always suitable to detect the hierarchical structure and the group composition of the system. When the correlation matrix is block diagonal, the eigenvalue spectrum has a number of large eigenvalues equal to the number of groups. Moreover, each corresponding eigenvector  has non vanishing components only for the elements of a specific group. In this case spectral analysis directly allows to identify a partition of the variables. However these properties are no more true when the system is intrinsically hierarchically organized. In fact, the number of large eigenvalues can be different from the number of groups and the eigenvectors of the correlation matrix associated with large eigenvalues have in general all non vanishing components, i.e. large eigenvalues cannot be associated with  specific groups of variables. 
In the supplementary material of this paper we describe in detail two simple HNFMs for which the direct eigenvectors' analysis fails in identifying the groups and in unveiling the hierarchical structure of the system. This result suggests that it is not possible to associate the largest eigenvalues neither with specific groups of elements controlled by the same factors nor with a common behavior mode governing all elements of the system when the nested nature of groups of elements is significant.
To make a specific example, consider a financial market. It has been recently suggested that there is a one to one association between the largest eigenvalues of the correlation matrix of stock returns with the global market behavior \cite{Laloux1999,Plerou1999,Gopikrishnan2001} or specific economic sectors \cite{Plerou1999,Gopikrishnan2001}. If financial market is hierarchically organized (as proposed below) this association 
might be less straightforward than originally thought (see also Ref.~\cite{krakow}). In conclusion, basic spectral methods, such as principal component analysis, could be unable to  fully describe the nested nature of hierarchical complex systems. For these cases our HNFM guarantees a proper hierarchical description of the elements of the investigated  complex system.  
To the best of our knowledge, HNFM is the first model based on empirical data in which both the dependency of variables from factors is nested and the factors are independent one of each other. This choice allows to consider the hierarchical clustering procedure from a perspective which is different from the one which is commonly adopted. Hierarchical clustering is not a tool which is only used to extract a partition of the elements but rather it is a tool that can also be used to associate a set of factors directly controlled by the genealogy of the element in the considered dendrogram with each element of the system. We believe this approach is useful in all the cases where a partition of the complex system is not straightforwardly feasible due to the fact that the system is clearly characterized by nested levels of hierarchies.

Eq. (\ref{model}) defines a HNFM of $N-1$ factors obtained from a dendrogram of $N$ elements. In general the number of factors determining the dynamics of the system can be significantly smaller than $N-1$. Moreover, several studies based on random matrix theory \cite{Laloux1999,Plerou1999} have shown that a correlation coefficient matrix obtained from a finite multivariate time series has associated an unavoidable statistical uncertainty that does not allow to discriminate between real and spurious factors. 
To overcome this problem, we propose here a method devised to select the HNFM characterized by the largest number of factors (although in any case less than $N$) compatible with a predefined threshold of statistical reliability of retained factors.
Our method exploits the technique of non parametric bootstrap \cite{Efron79} which is widely used in phylogenetic analysis. 

The method is illustrated below after we briefly sketch the procedure used to associate a bootstrap value with each internal node of a dendrogram. 
Consider a system of $N$ time series of length $T$ and suppose to collect data in a matrix ${\bf X}$ with $N$ columns and $T$ rows.  A bootstrap data matrix ${\bf X}^*$ is formed by randomly sampling $T$ rows from the original data matrix ${\bf X}$ allowing multiple sampling of the same row. For each replica ${\bf X}^*$, the associated correlation matrix ${\bf C}^*$ is evaluated and a dendrogram is constructed by hierarchical clustering. Some large number (typically 1000) of independent bootstrap replicas is generated and for each internal node of the original data dendrogram we compute the fraction of bootstrap replicas (commonly referred to as bootstrap value) preserving the internal node in the dendrogram. Given an internal node $\alpha_k$ of the original dendrogram we say that a bootstrap replica is preserving that node if and only if a node $\alpha_h^*$ in the replica dendrogram exists and identifies a branch characterized by the same leaves identified by $\alpha_k$ in the original dendrogram. For instance, we say that the node $\alpha_3$ of the dendrogram in Fig. \ref{dendro} is preserved in some replica dendrogram $D^*$ if and only if a node of $D^*$ exists such that it belongs to the genealogy of all and only the leaves 5, 6, 7, 8, 9 and 10.
The bootstrap technique allows to associate a bootstrap value with each internal node of a dendrogram. Because of the one to one relation between nodes in the dendrogram and factors in the HNFM, the bootstrap value associated with a certain node of the dendrogram is associated also with the corresponding factor in the HNFM. 

\begin{figure}
\includegraphics[width=0.49\textwidth]{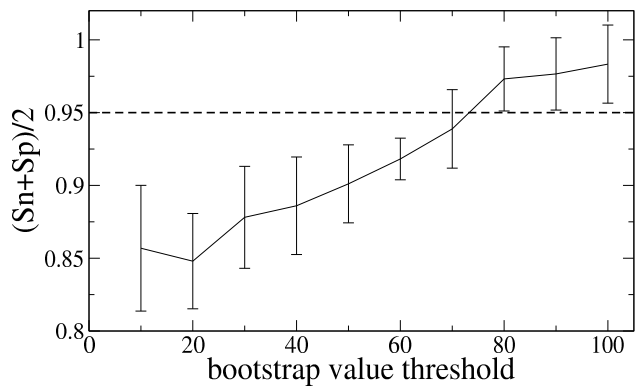}
\caption{${\cal R}=(Sn+Sp)/2$ as a function of the bootstrap value threshold. The error bar is one standard deviation. The dashed line indicates the chosen threshold of statistical reliability.} 
\label{reliability}
\end{figure}

Since the bootstrap value is a measure of the node's (factor's) reliability, we propose to remove those nodes (factors) with bootstrap value smaller than a given threshold $b$. This is done by merging each node with a bootstrap value smaller than $b$ with its first ancestor node in the path to the root having a bootstrap value greater than $b$ and then by constructing the HNFM associated with this reduced dendrogram. The question is how to select a suitable threshold $b$. The bootstrap value of a certain node (factor) cannot be straightforwardly intended as the probability that the node (factor) belongs to the true and unknown hierarchy (model) of the system.
For example, in phylogenetic analysis it has been shown \cite{Hillis93} that a bootstrap value of more than $70\%$ corresponds to a probability of more than $95\%$ that the true phylogeny has been found. By adapting the technique of Hillis and Bull \cite{Hillis93}, we do not choose {\it a priori} the value of $b$ but we infer  a suitable value of the threshold from the data in a self consistent way.  Specifically, we choose a certain number of bootstrap value thresholds $b_i$, e.g. $b_i=(i \times 10)\%, i\in\{0,1,...,10\}$. For each value of $i$, we remove internal nodes from the dendrogram according to $b_i$ obtaining a reduced dendrogram $D_i$ and a corresponding HNFM labeled HNFM$_i$. For each value of $i$, we perform $n$ simulations of data according to HNFM$_i$ and we label ${\bf X}_{ik}$ with $k\in\{1,...,n\}$  the data matrix of each simulation \cite{notedist}. To each ${\bf X}_{ik}$ we apply the clustering algorithm and the bootstrap node removal with the same threshold $b_i$ obtaining a reduced dendrogam $D_{ik}$. 
In order to compare the reduced dendrogram $D_i$ of the original data with the reduced dendrogram $D_{ik}$ of the data simulation we measure the sensitivity $Sn$ and specificity $Sp$ (see, for instance, \cite{Baxevanis}). In our case, the sensitivity $Sn_{ik}$ is the number of nodes in $D_i$ that are preserved in the reduced dendrogram $D_{ik}$ divided by the total number of nodes in the reduced dendrogram $D_i$. The specificity $Sp_{ik}$ is the number of nodes in $D_i$ that are preserved in the reduced dendrogram $D_{ik}$ divided by the total number of nodes in the reduced dendrogram $D_{ik}$. By averaging $Sn_{ik}$ and $Sp_{ik}$ over the $n$ different simulations we obtain the sensitivity $Sn_{i}$ and specificity $Sp_{i}$ of the node reduction associated with each bootstrap value threshold $b_i$. Finally, we obtain a measure of reliability of the dendrogram $D_i$ and of the corresponding HNFM$_i$ obtained for each bootstrap value threshold $b_i$, by averaging specificity and sensitivity ${\cal R}_i=(Sn_i+Sp_i)/2$ \cite{Baxevanis}. Note that we have defined sensitivity and specificity in terms of the nodes of the dendrogram $D_i$ which are preserved in $D_{ik}$. In an equivalent way  $Sn_{ik}$ and $Sp_{ik}$ can be defined in terms of the preserved factors in the corresponding models, HNFM$_i$ and HNFM$_{ik}$, i.e. the factors which determine the dynamics of exactly the same variables in both models. ${\cal R}_i$ can be interpreted as the probability averaged over all factors of the HNFM$_i$ that a HNFM$_{i k}$ contains a factor which is also present in the HNFM$_i$. Removing factors from the HNFM reduces the quantity of the empirical variance explained by the model. Therefore a satisfying bootstrap value threshold corresponds to the minimal value of $b_i$ such that ${\cal R}_i$ is larger than some standard threshold of reliability, e.g.  $95\%$ or $99\%$. In the example shown in Fig. \ref{reliability} (discussed below)  ${\cal R}_i>95\%$ for  $b_i\ge 80\%$. 
Finally, it should be noted that no assumption about the data distribution is needed to implement the method. \\
We have concluded above that the matrix ${\bf C^<}$ obtained by applying some hierarchical clustering technique to a correlation matrix is positive definite, provided that its elements are non negative numbers. Of course the same holds true for the matrix of the HNMF reduced according to the described bootstrap technique. 

\begin{figure}
\includegraphics[width=0.49\textwidth]{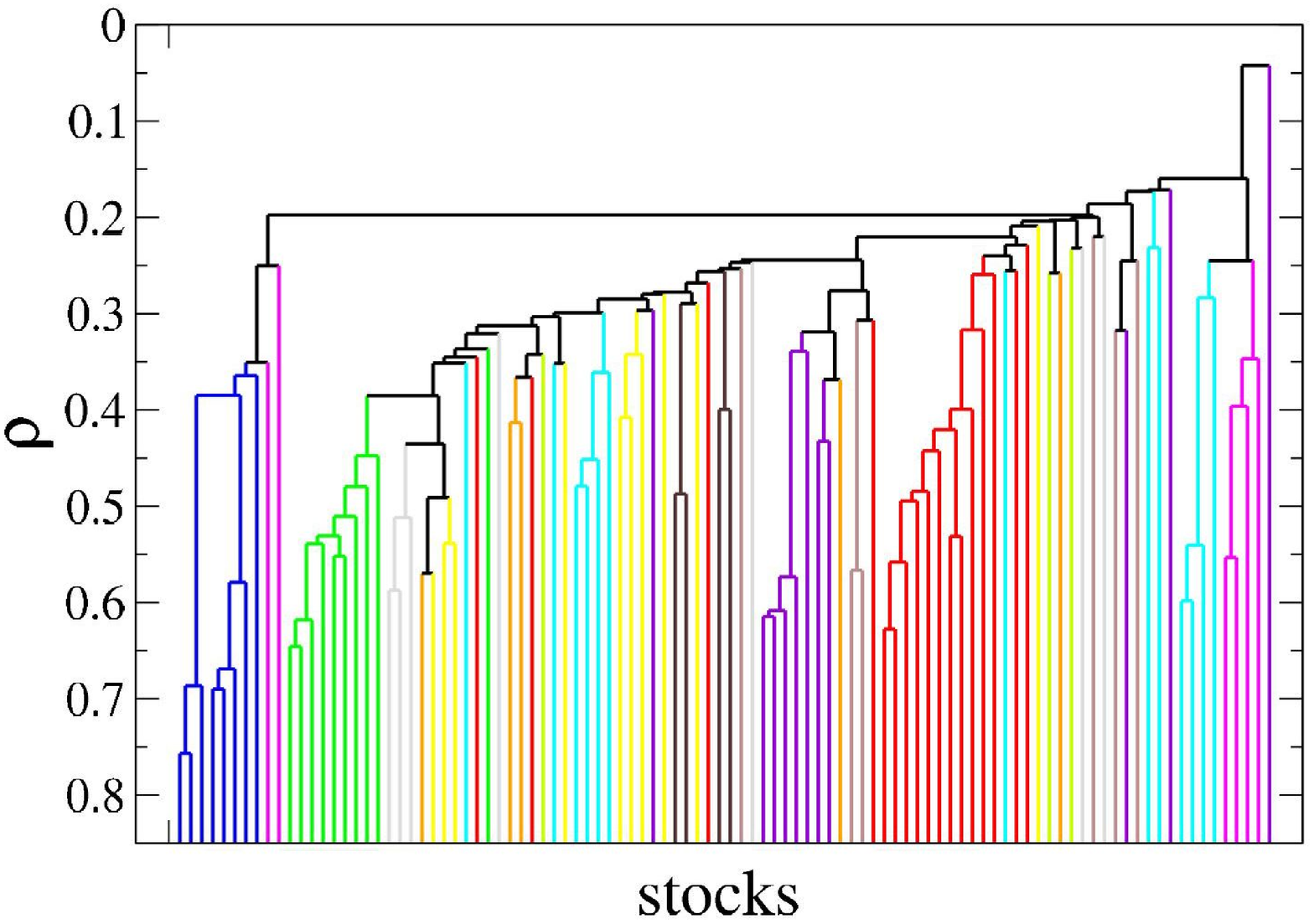}
\caption{Dendrogram of the set of daily equity returns of $100$ highly capitalized stocks traded at the NYSE during the period 1995-1998 obtained by applying the ALCA to the correlation matrix. Colors are chosen according to the stock economic sector.  Specifically these sectors are Basic Materials (violet), Consumer Cyclical (tan), Consumer Non Cyclical (yellow), Energy (blue), Services (cyan), Financial (green), Healthcare (gray), Technology (red), Utilities (magenta), Transportation (brown), Conglomerates (orange) and Capital Goods (light green).} 
\label{dendroData}
\end{figure}

As an application of the described technique to real data we examine a system monitored by recording the set of daily equity returns of $N=100$ highly capitalized stocks traded at the New York Stock Exchange (NYSE) during the period 1995-1998 ($T=1011$). We apply the ALCA to the correlation matrix of the system and we obtain the dendrogram shown in Fig. \ref{dendroData}. The dendrogram has $N-1=99$ nodes. The statistical reliability of these nodes is different from node to node due to metric and topological characteristics. The metric properties depend on the correlation coefficient values whereas the topological characteristics are depending on the ranking of these values and therefore on the complexity and number of hierarchies of the system. 

\begin{figure}[b,t]
\includegraphics[width=0.49\textwidth]{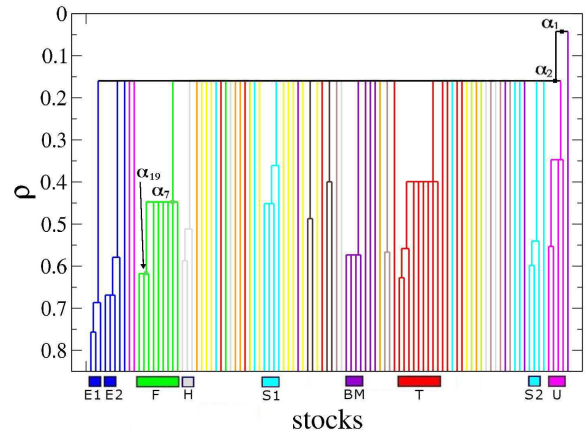}
\caption{Dendrogram with 23 internal nodes obtained by node reduction of the ALCA dendrogram (shown in Fig.~\ref{dendroData}) of $100$ stock daily returns traded at the NYSE during the period 1995-1998. Rectangles at the bottom are indicating 9 clusters and symbols label the classification of stocks in terms of economic sectors or sub-sectors according to the classification of Forbes' magazine. Specifically, E1 is the sub-sector of {\it Oil well services and equipment} and E2 is the sub-sector of {\it Oil and gas integrated}. Both E1 and E2 belong to the economic sector of {\it Energy}; T and F are indicating the economic sectors of {\it Technology}  and {\it Financial} respectively; H indicates the sub-sector  {\it Major drugs} of the economic sector {\it Healthcare}; BM indicates a cluster of stocks within the {\it Basic Material} economic sector. S1 and S2 indicate the two sub-sectors of {\it Communication services} and {\it Retail} of the sector of {\it Services} respectively. Finally, U is representing the sub-sector {\it Electric utilities} of the sector {\it Utilities}. Colors are chosen according to the stock economic sector as described in the caption of Fig.~\ref{dendroData} and the ordering of the stocks is the same as in Fig.~\ref{dendroData}. The labeled internal nodes are discussed in the text. In the figure we do not comment on clusters composed by only two leaves.} 
\label{nodered}
\end{figure}

We use the bootstrap technique described above, in order to evaluate the statistical reliability of each node and to simplify the description in terms of a HNFM. In particular, we select the minimal bootstrap value threshold that guarantees a value of ${\cal R}_i>95\%$. We accordingly reduce the number of factors of the corresponding HNFM. In our investigation, the number of bootstrap replicas is $1000$ and the number of simulations performed for each bootstrap value threshold is $n=20$. Simulated time series have been constructed by using original data. 
In Fig. \ref{reliability} we plot ${\cal R}_i$ as a function of the bootstrap value threshold. A direct inspection shows that the bootstrap value threshold $b_i=80\%$ guarantees that  ${\cal R}_i > 95\%$. The corresponding reduced dendrogram has 23 nodes and it is reported in Fig. \ref{nodered}. 

Let us first comment the properties of the reduced HNFM. 
In the figure we observe several clusters and sub-clusters. 
As already noticed in previous studies \cite{Mantegna1999, Tumminello05, Giada01, Gopikrishnan2001,krakow}, the detected clusters and sub-clusters are overlapping in part with economic classification such as the one provided by the Forbes magazine. This can be seen in Fig. \ref{dendroData} and \ref{nodered} where we use this classification to characterize with a specific color each stock. For example, financial firms are represented in Fig. \ref{dendroData} and \ref{nodered} as green lines in the hierarchical tree. 
One prominent example is the group of financial stocks. For illustrative purposes, let us consider the equations of the financial elements of the reduced HNFM. The first three stocks from left to right of the group labeled as F in Fig. \ref{nodered} are described by the equation $x_i^F(t)=\gamma_{\alpha_{19}} f^{(\alpha_{19})}(t)+\gamma_{\alpha_{7}} f^{(\alpha_{7})}(t)+\sum_{h=1}^{2}{\gamma_{\alpha_{h}} f^{(\alpha_h)}(t)} +\eta_F \epsilon_i(t)$. The factor $f^{(\alpha_1)}(t)$ is common to all stocks and  $f^{(\alpha_2)}(t)$ is common to all stocks except one, with tick symbol HM, which is a gold company. 
The factor $f^{(\alpha_{19})}(t)$ is specific to these financial stocks (their tick symbols are BAC, JPM and MER). The other six financial stocks also belonging to the same group (indicated by the tick symbols AGC, AIG, AXP, ONE, WFC and USB) are described by the equations  
$x_i^F(t)=\gamma_{\alpha_{7}} f^{(\alpha_{7})}(t)+\sum_{h=1}^{2}{\gamma_{\alpha_{h}} f^{(\alpha_h)}(t)} +\eta_F \epsilon_i(t)$.
In this last case only the $f^{(\alpha_7)}(t)$ factor is present in addition to the $f^{(\alpha_1)}(t)$ and $f^{(\alpha_2)}(t)$ factors common to all financial stocks. Since the factor $f^{(\alpha_7)}(t)$ is determining the dynamics of only financial stocks (9 out of 10 in the investigated sample), it is natural to consider $f^{(\alpha_7)}(t)$ as a factor characterizing financial stocks whereas $f^{(\alpha_{19})}(t)$ is an additional factor further characterizing only the three stocks BAC, JPM and MER. A similar organization in nested clusters is observed in all the groups detected by the reduced HNFM. The number of factors characterizing the various stocks is ranging from one to five. It is worth noting that each group of stocks, which are sharing at least 3 factors, is homogeneous with respect to the economic sector.

It is also worth to compare Fig. \ref{dendroData} and \ref{nodered}. The comparison shows that the self-consistent reduction of the number of factors allows a robust statistical validation of the groups that are detected from the data analysis. Only the information which is statistically robust at the $95\%$ level is retained in the reduced HNFM. For example, the energy cluster observed in Fig. \ref{dendroData} (blue lines in the figure) is not robust at the selected confidence level, whereas the two clusters indicated as E1 and E2 in Fig. \ref{nodered}, corresponding to the sub-sectors {\it Oil well services and equipment} and {\it Oil and gas integrated},  are robust. In Fig. \ref{nodered} all the detected clusters of more than $2$ elements and consistent with the Forbes classification are indicated by rectangles at the bottom of the figure. The economic characterization of clusters is discussed in the figure caption. 

In summary, we have introduced a method for associating a hierarchical factor model with a multivariate data set. The factor model is retaining all the information about hierarchies extracted from data by a hierarchical clustering procedure. We have also provided a bootstrap based procedure to obtain the HNFM with the largest number of factors compatible with a predefined threshold of their statistical reliability. This procedure selects in a self-consistent way the optimal bootstrap threshold for the considered set of  data. 
We have also shown that the similarity matrix ${\bf C^<}$, which is the output of hierarchical clustering procedures, is the proper correlation matrix of our model and therefore it is positive definite. 
Finally, we have used the HNFM to model a financial system of 100 highly capitalized stocks traded at NYSE.
This empirical analysis has shown the ability of HNFM in the modeling of a complex system characterized by nested levels of hierarchies inferred from data.

\acknowledgments{We acknowledge partial support from MIUR research project ``Dinamica di altissima frequenza nei mercati finanziari'', MIUR-FIRB research project RBNE01CW3M and NEST-DYSONET 12911 EU project.}

\section{Appendix}\nonumber

In the present supplementary material we introduce two simple time series models and we compare how straightforward spectral methods and hierarchical methods are able to unveil the hierarchical properties of the models. The two models are  hierarchically nested factor models.

As a first example, we consider a model (already introduced in \cite{Lillo05}) in which the $N$ variables follow a common factor $f_0(t)$ and two other factors $f_1(t)$ and $f_2(t)$ which are affecting two distinct groups of $n_1$ and $n_2=N-n_1$ elements respectively. 
The equations of the model are
\begin{eqnarray}
x_i(t)=\gamma_0 f_0(t)+\gamma_1 f_1(t)+\eta_1 \epsilon_i(t),  \forall  i \le n_1 \,\,\,\,\,\,\,\,\,\,\,\,\,\,\,\,\,\,\,\,\,\, \nonumber \\ 
x_i(t)=\gamma_0 f_0(t)+\gamma_2 f_2(t)+\eta_2 \epsilon_i(t),  \forall i: n_1< i \le N,  \label{2layer}
\end{eqnarray}
where $\gamma_0$, $\gamma_1$, $\gamma_2$ and $\eta_i$ $(i=1,2)$ are parameters. In this equation the factors $f_i(t)$ and the terms $\epsilon_i(t)$ are independent noise terms with zero mean and unit variance. We consider again variables $x_i$ with zero mean and unit variance without loss of generality. This choice fixes the value of $\eta_i$. We set  $\rho_{\alpha_1}=\gamma_0^2$, $\rho_{\alpha_2}=\gamma_0^2+\gamma_2^2$ and $\rho_{\alpha_3}=\gamma_0^2+\gamma_1^2$.
The eigenvalue spectrum of the correlation coefficient matrix of this model has two large eigenvalues given by  $\lambda^{\pm}=[2+q_{+} \pm  (q_{-}^2+4n_1n_2\rho_{\alpha_1}^2)^{1/2}]/2$,  where $q_{\pm}=(n_1-1)\rho_{\alpha_3}\pm(n_2-1)\rho_{\alpha_2}$,  $n_1-1$ eigenvalues equal to $1-\rho_{\alpha_3}$ and $n_2-1$ eigenvalues equal to $1-\rho_{\alpha_2}$. 
Thus, despite the fact that the original factor model of Eq.~(\ref{2layer}) has three uncorrelated factors $f_i(t)$, $(i=0,1,2)$, the spectrum has only two large eigenvalues. One could be tempted to interpret these large eigenvalues and the corresponding eigenvectors as describing the collective dynamics or the dynamics of the two groups. By analyzing the eigenvectors, it can be seen that this is not the case. The eigenvectors of the two largest eigenvalues have infra-group degenerate components and neither the first nor the second eigenvector is in general proportional to the vector $\{1,1,...,1\}$ representing the common behavior driven by the factor $f_0(t)$. 
Similarly, when one attempts to associate the first two eigenvectors with the two groups, one is faced with the fact that the first two eigenvectors have all non vanishing components. Our model indicates that the association between eigenvectors and factors is correct only in the limit when the system can be divided in groups of variables and each group is driven only by one factor. The generalization of the model to the case of heterogeneous $\gamma$ parameters and/or the finiteness of empirical time series makes even more involved the task of associating factors with eigenvectors when the correlation matrix of the model has hierarchical features. 
On the other hand, by applying a hierarchical clustering procedure, e.g. single linkage, average linkage and complete linkage, to the correlation matrix of the model of Eq.~(\ref{2layer}) one obtains the hierarchical tree of Fig.~\ref{dendro}A. The corresponding HNFM coincides with the model of Eq.~(\ref{2layer}). We have verified that by applying the bootstrap method we have introduced in our paper, we obtain back the HNFM of Eq.~(\ref{2layer}) also when we take into account the role of a finite number of records of multivariate time series (in our simulations we set $T=1011$ and $N=100$). Moreover, simulations have been performed by assuming the variables $x_i(t)$  being either Gaussian distributed or Student-t distributed with $4$ degrees of freedom. In both cases the recovered HNFM is the same and coincides with the model of Eq.~(\ref{2layer}). The threshold of reliability used to reduce the number of factors in the HNFM is ${\cal R}=95\%$ and the hierarchical clustering algorithm used is the Average Linkage Cluster Analysis (ALCA). This result shows that our method based on a hierarchical clustering procedure is able to recover the structure of the HNFM whereas basic spectral methods such as, for example, the principal component analysis are unable to uncover it. More specialized spectral methods, such as the varimax and the promax (or oblique rotation) methods \cite{Rencher}, are in most cases also unable to transform the eigenvectors associated with the two large eigenvalues of the model of Eq. (\ref{2layer}) in such a way that each eigenvector has non-vanishing components only for the variables belonging to one of the two groups. 

The second example we wish to consider is again a 3-factor model but with a completely nested structure. The equations of the model are:
%\,\,\,\,\,\,\,\,\,\,\,\,\,\,\,\,\,\,\,
\begin{eqnarray}
x_i(t)=\gamma_0 f_0(t)+\gamma_1 f_1(t)+\gamma_2 f_2(t)+\eta_1 \epsilon_i(t),  \forall  i \le n, \nonumber \\ 
x_i(t)=\gamma_0 f_0(t)+\gamma_2 f_2(t)+\eta_2 \epsilon_i(t),  \forall i: n< i \le 2 n, \,\,\,\, \nonumber \\
x_i(t)=\gamma_0 f_0(t)+\eta_3 \epsilon_i(t),  \forall i: 2 n< i \le 3 n=N \,\,\,\,\,\,\,\,\,\,\,\,\,\, \label{3layer}
\end{eqnarray}
and, as in the previous case, we consider random variables with zero mean and unit variance. The dendrogram associated with this model is shown in Fig.~1B. The eigenvalue spectrum of the correlation matrix has $3$ large eigenvalues and $3$ small eigenvalues each one with degeneracy $n-1$. The most general case is analytically solvable but the eigenvalues and eigenvectors cannot be expressed in a compact way. Thus here we set  $\gamma_0=\gamma_1=\sqrt{\rho}$ and $\gamma_2=\sqrt{2 \rho}$. With these simplifying parameters, the model of Eq. (\ref{3layer}) is depending only on the parameters $n$ and $\rho$.   
The space described by the eigenvectors of the 3 largest eigenvalues is the space of vectors ${\bf z}=\{z_1=u,...,z_n=u,$ $z_{n+1}=v,...,z_{2 n}=v,z_{2 n+1}=w,...,z_{N}=w\}$, i.e. the space of vectors with infra-group degenerate components. When $n \rho\gg 1$, the first 3 eigenvalues are $\lambda_1\cong(3+\sqrt{7}) n \rho$, $\lambda_2\cong n \rho$ and $\lambda_3\cong(3-\sqrt{7}) n \rho$. Since the components of the corresponding eigenvectors are defined  only in terms of $u$, $v$ and $w$ we represent  eigenvectors as characterized by 3 parameters by using the formalism ${\bf s}=\{u,v,w\}$. It results that the non normalized eigenvectors are ${\bf s_1}=\{8+3 \sqrt{7},5+2 \sqrt{7},3+\sqrt{7}\}$, ${\bf s_2}=\{-1,1,1\}$ and ${\bf s_3}=\{3 \sqrt{7}-8,2 \sqrt{7}-5,\sqrt{7}-3\}$. This result implies that also in this case the first 3 eigenvalues are associated with eigenvectors with degenerate non vanishing infra-group components. Moreover, none of these eigenvectors is proportional to the vector $\{1,1,...,1\}$ representing the common behavior driven by the factor $f_0(t)$. On the other hand, by applying the ALCA to the correlation matrix of the model of Eq.~(\ref{3layer}) one obtains the dendrogram of Fig 1B. The HNFM corresponding to this dendrogram coincides with the model of Eq.~(\ref{3layer}). 

\begin{figure}
\includegraphics[width=0.49\textwidth]{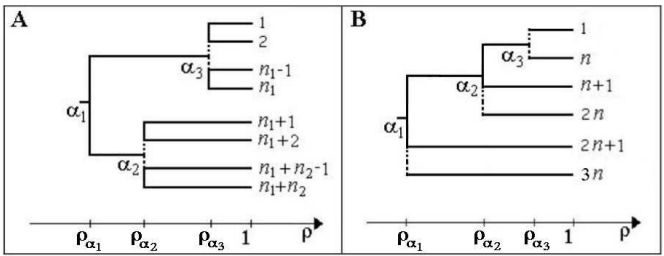}
\caption{ A) Dendrogram associated with the model of Eq. (1). B) Dendrogram associated with the model of Eq. (2)} 
\label{dendro}
\end{figure}

In summary this two examples of HNFM show that it is not always possible to associate the largest eigenvalues of the correlation matrix neither with specific groups of elements nor with all elements. It is also to notice that in the first example we have found 2 large eigenvalues in a system driven by 3 factors whereas in the second case we have observed 3 large eigenvalues for a model with 3 factors. This means that there is no direct relation between the number of factors in the HNFM and the number of large eigenvalues of the corresponding matrix ${\bf C^<}$.    These results indicate that standard spectral methods are not always suitable for the analysis of systems in which hierarchies are present.

%
% ----------------------------------------------------------------


\begin{thebibliography}{99}
\bibitem{note1}
Our method cannot in general be applied to clustering methods producing reversals in dendrograms, like in centroid methods.
\bibitem{note2}
When negative correlations are present among some pairs of variables, a suitable linear transformation of the similarity measure, which is not altering the dendrogram structure, can be defined.
\bibitem{notedist} 
The distribution of the factors and of the noise terms can be inferred from the data or one can use the investigated data set assuming it as representative of the underlying distribution.
\end{thebibliography}
\end{document}